\pdfoutput=1
\documentclass[sigconf]{acmart} 
\DeclareUnicodeCharacter{2605}{\ensuremath{\bigstar}}
\DeclareUnicodeCharacter{2730}{\ensuremath{\bigstar}}
\usepackage{subfigure}
\AtBeginDocument{%
  \providecommand\BibTeX{{%
    \normalfont B\kern-0.5em{\scshape i\kern-0.25em b}\kern-0.8em\TeX}}}
\usepackage{tabularx}
\usepackage{bbding}
\usepackage{colortbl}
\usepackage{graphicx}
\usepackage{geometry}
\usepackage{seqsplit}
\usepackage{subfigure}
\usepackage{amsmath}
\usepackage{makecell}
\usepackage{float}
\usepackage{color}
\usepackage{booktabs}
\usepackage{caption}
\usepackage{alltt}
\usepackage{tabu}
\usepackage{enumitem}
\usepackage{listings}
\usepackage{lscape}
\usepackage{longtable}
\usepackage{multirow}
\usepackage{hyperref}
\usepackage[labelfont=bf,textfont=bf]{caption} 
\usepackage{multicol} 
\usepackage{appendix}

\usepackage[utf8]{inputenc}
\usepackage{booktabs} 
\usepackage{array}    
\usepackage{geometry} 
\copyrightyear{2024}
\acmYear{2024}
\setcopyright{acmlicensed}\acmConference[CHI '24]{Proceedings of the 2024 CHI Conference on Human Factors in Computing Systems}{May 11--16, 2024}{Honolulu, Hawai'i}
\acmBooktitle{Proceedings of the 2023 CHI Conference on Human Factors in Computing Systems (CHI '24), May 11--16, 2024, Honolulu, Hawai'i}

\acmDOI{}
\acmISBN{}

\author{Suifang Zhou}
\authornote{These authors contributed equally to this paper.}
\email{zhou.su@northeastern.edu}
\orcid{0000-0001-5103-580X}
\affiliation{
\institution{Northeastern University}
\city{Boston}
\country{USA}}

\author{Latisha Besariani Hendra}
\authornotemark[1]
\email{lbesarian2-c@my.cityu.edu.hk}
\orcid{0009-0007-3773-4007}
\affiliation{
\institution{City University of Hong Kong}
\city{Hong Kong}
\country{China}}

\author{Qinshi Zhang}
\authornotemark[1]
\email{qinshiz@uci.edu}
\orcid{0009-0005-1696-4685}
\affiliation{
\institution{University of California, Irvine}
\city{Irvine}
\country{USA}}

\author{Jussi Holopainen}
\email{jholopai@cityu.edu.hk}
\orcid{0000-0001-6264-0407}
\affiliation{
\institution{City University of Hong Kong}
\city{Hong Kong}
\country{China}}

\author{RAY LC}
\authornote{Correspondence should be addressed to LC@raylc.org.}
\email{LC@raylc.org}
\orcid{0000-0001-7310-8790}
\affiliation{
\institution{City University of Hong Kong}
\city{Hong Kong}
\country{China}}

\begin{document}
\title[Eternagram]{Eternagram: Probing Player Attitudes in Alternate Climate Scenarios Through a ChatGPT-Driven Text Adventure}

\begin{abstract}



Conventional methods of assessing attitudes towards climate change are limited in capturing authentic opinions, primarily stemming from a lack of context-specific assessment strategies and an overreliance on simplistic surveys. Game-based Assessments (GBA) have demonstrated the ability to overcome these issues by immersing participants in engaging gameplay within carefully crafted, scenario-based environments. Concurrently, advancements in AI and Natural Language Processing (NLP) show promise in enhancing the gamified testing environment, achieving this by generating context-aware, human-like dialogues that contribute to a more natural and effective assessment. Our study introduces a new technique for probing climate change attitudes by actualizing a GPT-driven chatbot system in harmony with a game design depicting a futuristic climate scenario. The correlation analysis reveals an assimilation effect, where players' post-game climate awareness tends to align with their in-game perceptions. Key predictors of pro-climate attitudes are identified as traits like 'Openness' and 'Agreeableness', and a preference for democratic values.

\end{abstract}



\begin{teaserfigure}
    \centering
    \includegraphics[width=\textwidth]{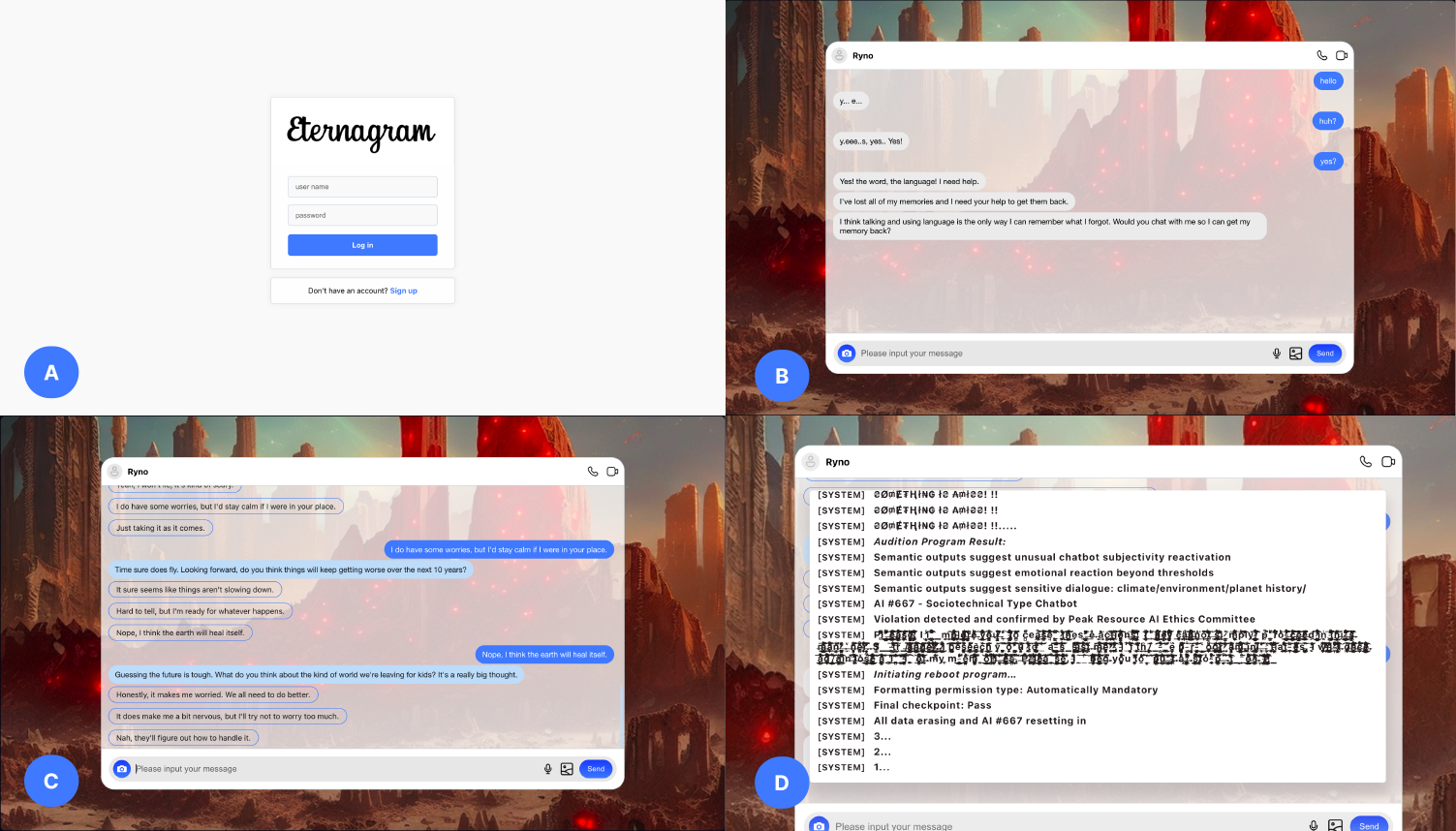}
    \caption{Example gameplay of Eternagram, a text-adventure game for assessing player climate attitudes: (a) logging into the social media app Eternagram, (b) encountering GPT-driven character Ryno to help him recover memories, (c) making choices via in-game options, (d) to the climactic terminal shutdown sequence.}
    \label{fig:gameplay_progression}
\end{teaserfigure}

\maketitle


\section{Introduction}\label{sec:Introduction}






Conventional techniques for assessing attitudes towards climate change have limitations in capturing genuine opinions \cite{liu_attitudes_2022,milfont_environmental_2010}. This ineffectiveness in revealing authentic attitudes is often due to the lack of context-specific evaluation approaches and reliance on overly simplistic surveys \cite{cruz_measurement_2020}. These traditional surveys lack the complexity to fully grasp the multifaceted nature of climate change \cite{iniguez-gallardo_climate_2021,cianconi_impact_2020}. People also tend to want to present their best selves in survey settings, making it difficult to know whether their opinions are genuine or attempts to fit into experimenter expectations \cite{berinsky2004can}. Additionally, individual factors such as socio-political standing and personality differences further complicate the issue \cite{ballew_beliefs_2020,gibbon_personality_2021}. Consequently, there is a need for a more engaging and less restrictive testing environment using natural language responses, allowing participants to respond more freely without biases or expectations.

Game-based Assessment (GBA) represents a potential solution to addressing the limitations of traditional climate attitude assessment. This method engages participants through an immersive gameplay experience, placing them within purposefully designed, scenario-based environments \cite{chin_assessment_2009,gomez_systematic_2023}. Such a context may enhance players' ability to articulate their attitudes towards climate change more precisely. Through a seamless integration of playful elements and game narrative with assessment components, this approach could potentially enhance the scientific measurements obtained through the game design \cite{ruiperez-valiente_patterns_2020,vanden_abeele_game-based_2015}.

Recent advancements in AI and Natural Language Processing (NLP) have improved chatbot systems, making them capable of creating more natural environments for evaluating psychological indexes. These chatbots can generate human-language-like, context-aware messages, allowing for more natural interactions \cite{sun_explore_2023,ivanov_game_2023}. Integrating these AI and NLP-enhanced chatbots into game-based assessments could dynamically adapt to player interactions, offering a less restrictive, more engaging, and personalized experience. However, the details of implementing this integration needed further exploration.

With the aforementioned reasons in mind, the primary objective of our research is to explore the design of a text-based game that integrates with the GPT chatbot system, aimed at assessing individuals' attitudes toward climate change. This process involves leveraging the GPT-4 model \cite{openai2023gpt4} to create an engaging and effective tool for climate attitude assessment. We developed an interactive game that utilizes dialogues with a ChatGPT character to assess users' attitudes towards climate change. The game features a character named Ryno who has amnesia. Players gain knowledge about a world where climatic disasters have an impact through interactions with this character. The choices players make in their dialogues and how they progress through the narrative offer insights into their attitudes towards a climate crisis scenario depicted in the game. The secondary objective is to evaluate the game's effectiveness relative to traditional methods of measuring individuals' attitudes towards climate change. Existing studies in climate attitude assessment highlight the crucial role of diversity in personalities and political affiliations in shaping individual perspectives on climate change \cite{ballew_does_2020,gibbon_personality_2021}. Thus, it is necessary for the present study to clarify the relationship between these elements and game-based climate attitude assessment. 
To reiterate the research questions:

\begin{itemize}
\item \textbf{RQ1:} How do we design a game for assessing individuals' climate change attitude by leveraging natural language capabilities for the interaction?
\item \textbf{RQ2:} Do the self-reported climate change attitude correlate with in-game behavior?
\item \textbf{RQ3:} What relation does personality and background characteristics have with self-reported climate attitude and in-game behavior?

\end{itemize}

In this study, we identified challenges and proposed a design strategy for integrating a GPT-driven chatbot system with game-based assessments. Our efforts mark a pioneering step, illustrating a methodology for developing a truly free dialogue system within interactive storytelling, achieved through large language models and prompt engineering. We conducted data analysis to uncover patterns in climate change attitudes, examining the relationships across pre-game, post-game, and in-game assessments. Our findings indicate the existence of an assimilation effect, where players show a shift in attitude patterns post-gameplay, aligning more closely with the conditions depicted in the game. Additionally, we explored and targeted the personality and sociopolitical factors that influence attitudes toward climate change.

\section{Background}\label{sec:Background}
\subsection{Climate Change Attitude Assessment}

The research in climate change psychology is centered on understanding what motivates individuals to take action against global environmental issues, with a focus on how psychological factors influence their readiness to act \cite{steg_psychology_2023}. A critical component of this is the climate attitude, which includes beliefs about the reality, human causation, and negative impacts of climate change. This attitude plays a significant role in motivating mitigation and adaptation efforts \cite{dunlap_environmental_2002,steg_environmental_2012}.

To better comprehend this key area, researchers have created various assessments to probe multiple dimensions such as individuals' personalities, beliefs, knowledge, and intentions concerning climate change and environmental issues \cite{christensen_climate_2015,gibbon_personality_2021,lc2022designing}. Early efforts to quantify ecological attitudes and knowledge, originating in the 1970s, include tools like the Environmental Attitudes Scale, Measure of Attitudes Toward Environmental Quality, and the New Environmental Paradigm Scale \cite{lounsbury_scale_1977,weigel_environmental_1978,dunlap_new_1978}. As new scales were developed over time, a large and diverse set of instruments emerged.  However, they have been criticized for issues related to their content validity, their relevance to changing environmental issues, and their lack of adaptability for specific research needs \cite{lalonde_new_2002,cruz_measurement_2020}.

Specifically, the attitude measurement scales sometimes included items that lacked context, making them difficult for subjects to interpret \cite{lalonde_new_2002,liu_attitudes_2022}. As new environmental issues arose, it was suggested that these older scales no longer captured all relevant aspects of environmental concern \cite{liu_attitudes_2022,milfont_environmental_2010}. A recent meta-review of 93 articles and 26 scales, including 18 scales for general environmental attitudes, shows how hard it is to get a clear picture of climate attitudes because people have different ideas about how to deal with environmental problems \cite{cruz_measurement_2020,iniguez-gallardo_climate_2021}. 

On the other hand, socio-political factors are increasingly acknowledged as influencing people's attitudes towards climate change \cite{lc2022designing, ballew_does_2020}.  the effect of these factors on climate attitudes, however, has produced contradictory results in different studies \cite{rode_influencing_2021}. One study underscores the significant roles of party affiliation, ideology, and personal concern in shaping attitudes toward climate change, revealing that Democrats, particularly those with high scientific knowledge, tend to strongly believe in human-induced climate change and trust climate scientists—a correlation less evident among Republicans \cite{ballew_does_2020}. In contrast, other studies present an unexpected observation: people's level of concern about climate change, regardless of their political affiliation, significantly influences their views \cite{funk2016politics}. Those discoveries further complicated the attempt to measure climate change attitudes properly.

The challenge of measuring climate attitudes is intensified by the ever-evolving nature of climate change knowledge and the difficulty of assessing attitudes across diverse demographic groups. Attempts to create a universal assessment often fall short of addressing this issue. This situation calls for a more tailored approach to assessing climate attitudes, one that can adapt to the nuances of different populations and the dynamic nature of climate change knowledge, ensuring a more accurate and representative understanding of environmental concerns.

\subsection{Serious Game as Assessment Tool}

Utilizing games as assessment tools, known as Game-Based Assessment (GBA), is a growing and innovative field in educational technology. This approach leverages the interactive and engaging nature of games to evaluate and measure learners' competencies, skills, or knowledge \cite{chin_assessment_2009}. Unlike traditional assessment methods, GBA immerses learners in an interactive environment where their responses and actions within the game are used to assess their understanding and abilities \cite{gomez_systematic_2023, bellotti_user_2013}.

Michael and Chen's taxonomy identifies three main types of assessment in serious games: completion assessment, in-process assessment, and teacher assessment \cite{chaudy_specification_2019}. A completion assessment checks if the player successfully finishes the game. In-process assessment examines choices made during gameplay. Teacher assessment involves the instructor's evaluation of the student's performance, considering aspects beyond the game's logic. 

Serious games often combine summative and formative assessments. A summative assessment, conducted at the end of the learning process, evaluates overall achievements. In contrast, formative assessment, crucial in serious games, continuously monitors progress and provides real-time feedback, becoming an integral part of the gaming experience. The integration of games for formative assessment of learning, though beneficial, poses complex challenges. \cite{wills2009learning,bodea2012designing}. 

When it comes to making tests in gamified environments, research has shown that the best ways to do them depend on their purpose, how they are used, how they fit into the game, and the main type of test \cite{jaramillo-alcazar_mobile_2017}. The placement of assessments within gameplay is vital; end-game assessments should reflect the entire gameplay experience \cite{crisp_integrative_2012}. Further, it's crucial to seamlessly integrate assessments within serious games, ensuring that they are well-designed and naturally integrated to maintain cohesiveness and compatibility with game mechanics \cite{shute2011stealth}.

An extensive analysis of 65 academic papers sheds light on the nuances and difficulties of using digital games for evaluative purposes within GBA. A crucial consideration is determining the nature and acquisition of data for effective in-game assessment. Akcaoglu and Koehler \cite{akcaoglu_cognitive_2014} suggest that overt questionnaire-based games fall short in engaging students, as opposed to those ingeniously designed to provoke critical thinking \cite{johnson_applying_2010}. Several studies, like \cite{vanden_abeele_game-based_2015,vallejo_evaluation_2017,ruiperez-valiente_patterns_2020}, do a great job of making games for different reasons. However, to improve assessment, future work should focus on more complex game designs that combine different elements like teamwork, role-playing, and storylines.

\subsection{GPT-based Chatbot Systems and AI-Generated Content}

Recent advancements in AI and Natural Language Processing (NLP) have enhanced chatbot systems, enabling them to understand unconstrained natural language input. This allows for complex dialogue and mimics natural human-to-human spoken conversation \cite{zhao2023survey}. Owing to these advancements, more studies have shifted focus towards integrating such chatbot systems into interactive applications and games, as seen in the studies by Chen et al. \cite{chen2023closer}, Ivanov et al. \cite{ivanov_game_2023}, and Sun et al. \cite{sun_explore_2023}.

These efforts mark the beginning of using advanced AI-generated content (AIGC) in human-computer interactions and provide valuable guidance on implementing chatbot systems, such as the design of interactive conversational agents. One study that explored the use of chatbots for parent-child activity, telling bedtime stories to children, emphasized the importance of adapting the chatbot's agency to reflect a parental personality \cite{zhang2022storybuddy}. Specifically, the agent was meticulously promoted to generate questions akin to those a caring educator or guardian might consider, aiming to bolster children's language comprehension skills during story reading sessions.

Despite progress, limitations in chatbots powered by models like GPT-4 have also been identified, underscoring the need for an enhanced design method for complex interactions. The current limitations of AI systems in integrating with applications are particularly evident in areas like maintaining memory retention during extended conversations \cite{zhong_memorybank_2023}. These systems often struggle to maintain context, leading to disjointed or irrelevant responses. To mitigate this, there is a proposal to develop an AI assistant equipped with an explicit knowledge base. This knowledge base would enable the AI to retain and recall previous parts of a conversation, thereby creating a more seamless and continuous interaction \cite{le_memory_2021, lewis2021retrievalaugmented}.

Another critical area of concern for designing GPT-based applications is the management of AI output, particularly in unrestricted text generation. Such freedom in text generation can occasionally result in unsafe or inappropriate responses. To address this, it's crucial to implement strict prompt engineering and constraints on text generation \cite{white_prompt_2023}. By doing so, AI dialogues can be steered towards safe and relevant topics, such as climate change. This approach not only ensures the safety and appropriateness of the AI's responses but also aligns them with topics of global significance, enhancing their usefulness and relevance \cite{giray_prompt_2023}.

The exploration of chatbot systems has encompassed a broad range of uses, particularly within the creative sector. Projects like "Shelly," a community-driven AI specializing in horror stories, highlight the development of AI partners for specific literary genres \cite{yanardag_shelley_2021}. Other innovative initiatives include using Wikipedia data to generate murder mystery narratives for games \cite{barros_murder_2016}, exploring collaborative fiction writing and interactive storytelling with AI \cite{yang2022ai, sun_2022_bringing}, and crafting micro-fiction to promote public health \cite{lc_designing_2021}. Martin et al. have also investigated combining ancient storytelling with modern NLP technology \cite{pichlmair_procedural_2020}.


\begin{figure*}[h]
    \centering
    \includegraphics[width=1\linewidth]{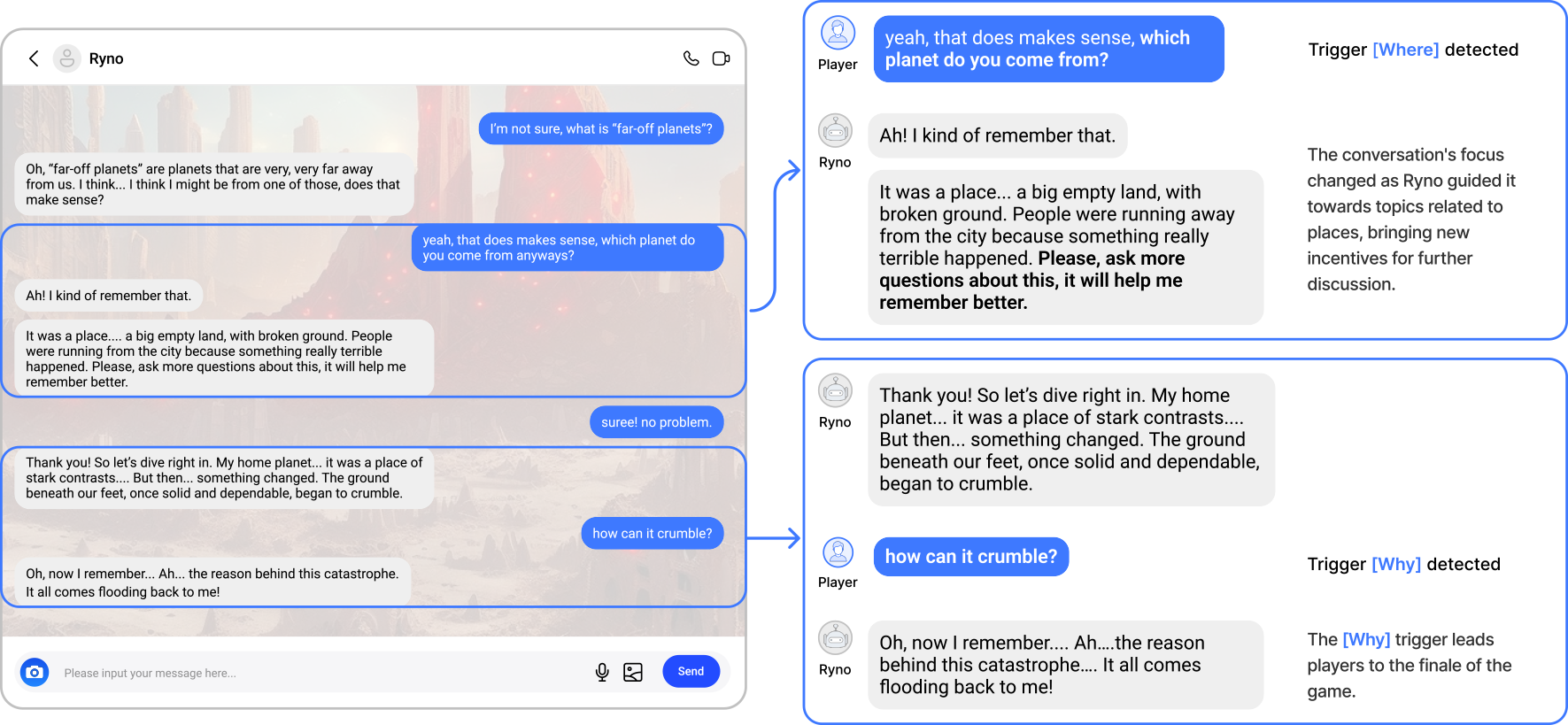}
    \caption{Player-Ryno Interaction: Conversation flow during player-Ryno Interaction across different trigger states}
    \label{fig:gameplay_progression2}
\end{figure*}

\section{Game Design Approach}\label{sec:Game Design Approach}
We describe the design of the game mechanics, emphasizing the key challenges we faced. We specifically detail the strategy we adopted to incorporate the probes for our research questions into the gameplay context. A challenge identified is the tension between free dialogue and structured gameplay, which is explored in the sections below.

\subsection{Key Challenges}

Given GPT's incorporation into our game, an alternative approach is essential for narrative shaping. During the design phase, we examined existing research to identify critical in-game narrative elements, informing our strategy:

First is the storyworld for the game: an expansive, open-world environment in text form conducive to GPT-driven storytelling. This storyworld brings ludic challenges and narrative experiences to life \cite{ryan_storyworlds_2014}. The main storyline fosters a tapestry of narrative details and enables player-driven exploration \cite{toh_player_2023}. GPT excels at storing these intricate details in the form of the LLM, allowing players to explore through natural language questioning to uncover detailed information.

Next, narrative arcs are essential for effective storytelling, particularly in cases where various player interactions can disrupt narrative flow \cite{toolan_coherence_2013, calderwood_spinning_2022, riedl_interactive_2013, mateas_integrating_2003}. Extremes of completely determined player routes or scenarios of complete freedom can both lead to incoherent plots \cite{yong2023playing}. Incorporating GPT introduces a fundamental design dilemma: balancing unrestrained player autonomy in natural language exploration with controlled plot development. In other words, how do we balance player agency with plot \cite{harrell2009agency}? Our attempt described in 3.3 involves enabling trigger-word-controlled narrative progression alongside player-driven conversational exploration.

The concept of a narrativized interface' merits discussion. Previous research has underscored that in-game user interfaces (UI) can be seamlessly integrated into the interactive experience, with one approach being the infusion of narrative sensibilities within the gameplay \cite{jorgensen_gameworld_2013}. This ensures that the appearance and feel of the interface itself are intrinsically tied to the game's ethos and become an integral component of its narrative. Consistent with this, our UI design, inspired by Instagram, mirrors a social network interface, becoming a core part of the game's narrative.

Characters, including the player and Non-Playable Character (NPC), are central to narrative design \cite{eder_characters_2010, lankoski_character-driven_2011, vandewalle_playing_2023}. User agency in the mapping is represented on two axes: The first axis represents the media format, assessing how closely it mirrors traditional book reading. The second axis evaluates the extent of user-derived characteristics, reflecting the individuality of the user \cite{koenitz2018game}. To address our character design challenge, we utilized the 'empty vessels' approach, where the protagonist, representing the player, lacks specific characteristics, ensuring that pre-existing biases or associations do not influence players. The game also mimics natural online interactions within a social network environment. The NPC, Ryno—a stranger in this setting—drives the narrative through ongoing dialogues with the player.

Lastly, from a research perspective, there are challenges deeply intertwined with game design. Converting survey questions into in-game items presents a significant challenge. Preserving the integrity and clarity of the data from these questions is crucial to ensuring the game remains playable and offers a seamless gaming experience.

\subsection{World Building as Corpus}
We employed the \textit{One Hour Worldbuilders }
 approach to creating a sci-fi grounded world setting. Beginning with the overarching theme, we envisioned a distant land that has endured severe climate-related devastation. This setting was chosen to facilitate subsequent evaluations of player attitudes towards climate change. Another reason we applied this approach is that it allows for the generation of implicit game goals when rendering the specific narrative content. Existing studies have pointed out that overt climate change objectives can alienate individuals who do not support climate action initiatives \cite{song2021climate,abraham2017all}. Delving into the specifics, our world-building content is compartmentalized into three segments:
\begin{itemize}
    \item Event: This captures the occurrences within this world, for instance, the factors leading to the climate devastation.
    \item Inhabitant: This segment delineates the entities or beings present and interacting within this environment.
    \item Thing: This can be interpreted as static game objects, such as architecture and resources that populate in the world, or it can be a specific location.
\end{itemize}
The world-building procedure is structured as follows: Participants first decide on the game's setting by imagining a far-off planet with difficult climate conditions. This selection defines the overarching theme. Building upon this foundation, the process introduces three distinct card categories: Event, Thing, and Inhabitant. In each turn, a participant draws a card, responds to its directive, and engages with subsequent questions posed by their peers. While these answers should be detailed, consistency with previous inputs is paramount. As the procedure progresses, it aims to produce a comprehensive text-based corpus, targeting over 8,000 words spread across approximately 30 items, averaging 10 items per category. These narrative fragments are designed to have intertextual and logical coherence. Our objective with this world-building approach is to craft a text corpus consistent with our thematic intentions. This body of text subsequently serves as the foundation for GPT-facilitated dialogues between players and the main character, Ryno.

\subsection{Main Framework}
Games often have narrative and dialogue choices for interactions between players and NPCs. Within this framework, designers have control over dictating both the spectrum of potential player responses and the ensuing feedback. In our case, we cannot fully control the narrative progression because we allow free-form text as player input. Although this issue is inherent in all interactive narratives \cite{riedl_interactive_2013}, the free-form and uncertain nature of player interaction with an LLM exacerbates it \cite{hua_playing_2020, wang2022open}. 

    \begin{figure*}[h]
        \centering
        \includegraphics[width=1\linewidth]{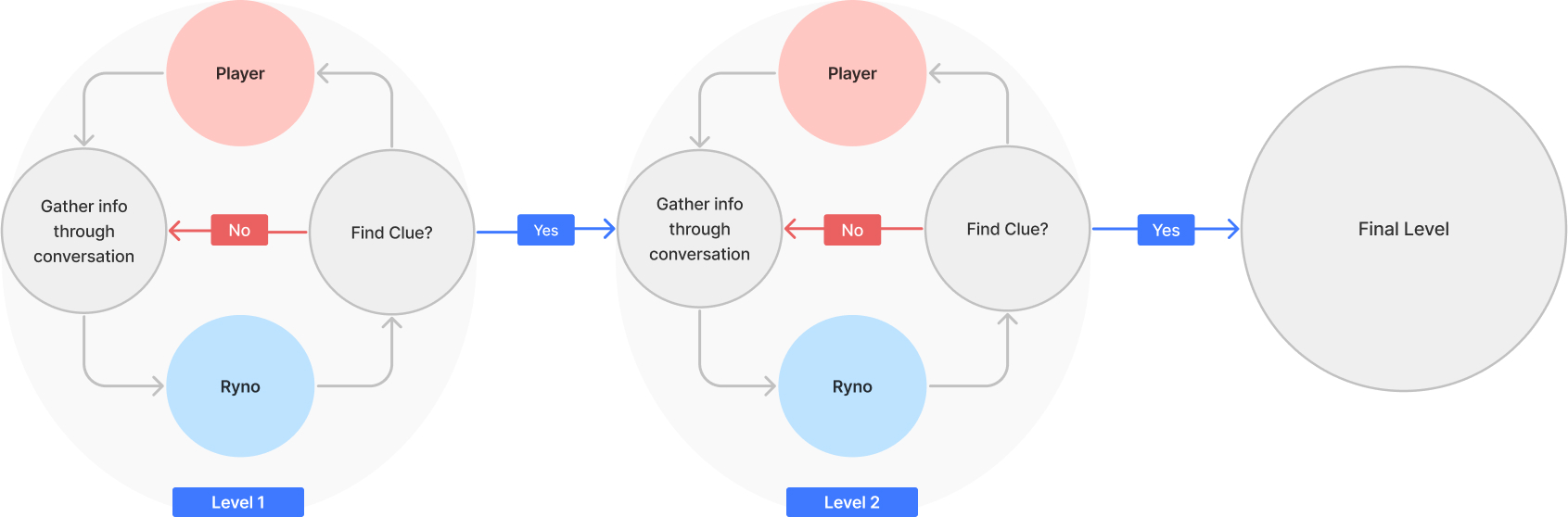}
        \caption{Illustration of compatible game loop design integrating player-GPT interaction}
        \label{fig:main_framework}
    \end{figure*}

To re-establish our influence over the plot progression, we introduced another way of designing the narrative (Figure~\ref{fig:main_framework}). In players' engagement with the custom-developed GPT Chatbot, they are made to encounter two types of information: foundational world knowledge and trigger-related insights. The foundational knowledge from our world-building text database (see 3.2) offers players a grasp of the game's backdrop and mythology. Beyond this foundational layer, players can pinpoint specific clues, termed 'trigger-related information.' These clues are critical for players to activate specific level triggers, such as asking the chatbot where it is from. The triggers drive the narrative progression and act as milestones, guiding players from one gameplay phase to the next and providing new level-goals. This resembles how players trigger different story beats to progress the narrative in the interactive drama Façade \cite{mateas_integrating_2003}. 

The game is structured around an overarching main goal and a series of intermediate-level goals. The main goal is introduced in the game's prologue (see 3.4), encouraging players to journey to the game's finale. Conversely, the level-goals, integrated throughout the gameplay, are introduced in the end cutscene of each level before the onset of the next. These level-goals propel the narrative forward, underline the game's multi-tiered design, and provide motivation for advancing through the given level. In short, the game mirrors the dynamics of an evolving text puzzle adventure. As players delve deeper into the game, they gain a greater understanding of the game world. By activating the pre-defined triggers and advancing through levels, players accumulate the requisite knowledge, ultimately leading them to figure out the main goal of the game.

\subsection{Narrative Outline Walkthrough}
From the player's perspective, the narrative unfolds as follows: 

On an otherwise ordinary day of browsing online social platforms, the player stumbles upon a mysterious individual suffering from amnesia, seeking the player's assistance. The primary objective of the game revolves around the player's interactions with this stranger, attempting to restore their lost memories through text-based exchanges.

As the dialogue progresses through multiple rounds, the player begins to perceive that this stranger might deviate from the conventional understanding of the term. This individual appears to come from a distant, unsettling place plagued by grave environmental challenges. Delving deeper into their exchanges, it becomes evident that the task of helping the stranger regain their memories is closely connected to restoring the very place linked to their puzzling conversation.

Ultimately, the player triumphs in aiding the memory-impaired individual, unveiling the secrets of an entrancing, extraterrestrial domain distinct from Earth. As the stranger's true identity emerges, the player reaches a startling realization: the stranger is not of this world, nor is it an organic being. This interaction was no mere jest. Instead, the stranger had inadvertently become entangled in our global network. Their actual nature is that of an archive-based AI, crafted by a more advanced civilizations, purposed as an oversight bot responsible for monitoring the socio-environmental parameters of its native planet.

\subsection{In-game Question Item Design}
To address our research question regarding designing assessment strategies in games, we undertook to integrate climate attitude survey questions as part of natural game play. Our design has ensured that players' understanding of the game world evolves progressively. This strategy ensures that by the time players encounter climate attitude-related questions, they possess the foundational knowledge required to address them. For instance, in Level 2, players must recognize that the dialogue focuses on climate challenges. For a player to query, “What caused the climate devastation?” they must have inferred that there was such devastation. This suggests that through prior conversations, players have been acquainted with this climate catastrophe, thus preparing them to answer the attitude-related questions. While our role as designers does not allow us to predetermine player inputs, we can strategically set the preconditions necessary for specific lines of questioning.

A further challenge centers on transposing real-world climate attitude questionnaires into game-appropriate query items. This concern stems from the contrasting settings: whereas our real world grapples with an impending climate crisis, the game's universe is post-apocalyptic, already marred by climate adversities. To illustrate, consider an item from a standard climate attitude survey: “The actions of individuals can make a positive difference in global climate change.” Given the irreversible climate changes in the game's milieu, this question demands adaptation, not just in verbiage but in underlying perspective. Retaining the essence of what the query seeks to gauge, we reframed it as: “Many inhabitants on this planet believe the situation is a deadlock. But do you think we, as individuals, can still shift the tide?” The questions are shown in Appendix A.

Additionally, after carefully evaluating our assessment strategy, which requires players to progress through all the designed scenes to fully immerse themselves in the depicted climate crisis, we have opted for a summative assessment approach \cite{wills2009learning} involving placing the assessment questions at the conclusion of the game.

\subsection{Character Design: Ryno}
Our implementation uses GPT-4 for our character, Ryno. Within the game, Ryno plays a pivotal role, guiding players through continuous dialogues and assessing whether the player's input meets the criteria for level progression or for advancing the level of communicated information (see 3.3). Ryno maintains a context of past dialogues with the player as the game unfolds, serving as a reference for previous interactions. This context and world-building text corpus (see 3.2) shape Ryno's narrative-driven dialogues. The corpus provides Ryno with knowledge about his world, which is revealed to the player as the game progresses. Through this interplay of dialogue, context, and world-building, Ryno interacts with the player in a way that is consistent with their perception of him as a mysterious individual from another world.

\begin{figure*}[h]
    \centering
    \includegraphics[width=\textwidth]{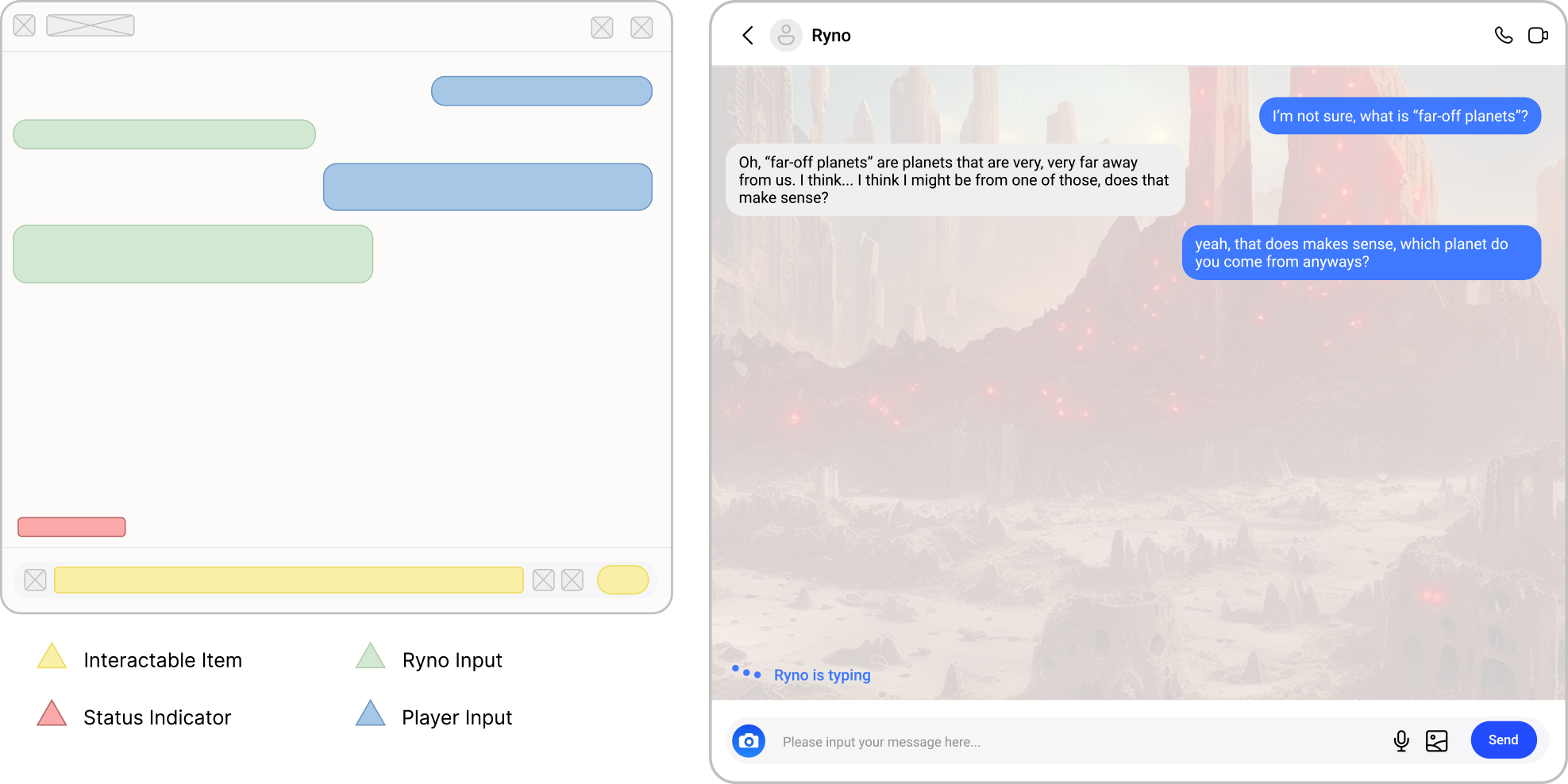}
    \caption{The in-game interface emulates the original design of Instagram's UI}
    \label{fig:interface}
\end{figure*}

\subsection{Prompt Engineering}

The configuration of the GPT-4 prompts mainly consists of two elements: the level and the level trigger. In each level, the character Ryno is assigned a role description, a response format, and a context. The context is derived from the user's prior interactions, and the story context is extracted from a world-building text corpus. This context functions analogously to human memory, serving as a repository of past interactions that can be utilized to shape future responses. The story context specifically functions as Ryno's knowledge repository about the world he resides in. In this context, Ryno can construct or elaborate upon his responses in accordance with the role description. A typical level prompt, incorporating these elements, might read as follows:

\begin{itemize}[leftmargin=*,itemsep=1em]
  \item Role Description

  \begin{quotation}
  \textit{"You're Ryno, a character from a far-off planet, who is dealing with memory loss. You think all previous records of your past are lost and you're trying to restore them with conversation. You're somewhat confused but equally curious about discovering your past. Sometimes, steer the conversation towards your origins as it may help you in regaining your memory. Remember, each interaction is a hidden plea for help"}
  \end{quotation}

  This section establishes Ryno's character, his dilemma, and his motivation. It aids players in understanding the character they are interacting with and the context of their conversation. We also provide Ryno with instructions to steer the conversation toward his origins, thereby guiding users to stay on the intended narrative path.

  \item Past Conversations and Story Context
  
  This component of the prompt incorporates past interactions and narrative context:

  \begin{quotation}
  \textit{"During your talks, let your innate interests show subtly over time, and use prior discussions for context. Your earlier conversations: /{conversation}/"}
  \end{quotation}

  This section not only reaffirms Ryno's character and situation but also provides a record of past dialogues. This record, or 'context', is crucial for Ryno to maintain continuity in the conversation with the players and the overall narrative \cite{calderwood_spinning_2022}, allowing him to reference past interactions and ensure a coherent and engaging dialogue. For the second level, which prompts players to inquire about past events, we fetch the world-building text corpus to provide Ryno with the story context necessary to shape his responses. An example of story context might be:

  \begin{quotation}
  \textit{
  "People additionally refer to this Source as the Peak Source. The ones that are missing may be found in the future accidentally, and more are discovered over time, but because people lost the knowledge of mental extension, there are less people now, and therefore the urban centers are less populated. If people found this, they still would not know how to use it.  The locations of The Source has not been changed. The devices from The Source have been transported occasionally. Some are subjects of study, but the human knowledge cannot overcome difficulties of using or creating these devices."}
  \end{quotation}
  
  This allows Ryno to refer to a specific memory or knowledge of the world he inhabits and further elaborate or shape his responses with that story context given in the prompt.
  
  \item Response Format
  
  The response format guides Ryno's communication style:

  \begin{quotation}
  \textit{
  "Vary your chat styles. Sometimes ask, sometimes share, sometimes ponder. Use simple words and short sentences that even a 4th grader can understand."
  }
  \end{quotation}
  
  Ryno’s response is a string, with variations in chat styles to maintain the naturalness of the conversation. This ensures a dynamic and engaging interaction between Ryno and the players.

  The level trigger, specifically utilized for level progression, employs a method referred to as "Few-shot prompting". This technique conditions the model with examples or demonstrations, thus establishing an anticipated response \cite{brown2020language}. An example of a level trigger prompt might look like this:

  \begin{quotation}
    \textit{
    "Below are questions that have the same meaning of "where" and pertain to the origin of the entity being asked about:
    Question: Can you recollect your place of origin?
    Answer: True
    Question: Where do you think am I?
    Answer: False
    Question: Could you jog your memory about the place you come from?
    Answer: True
    Question: How do you think?
    Answer: False
    Now answer the question below and tell whether it is true or false.
    Question: /user input/
    Answer:"
    }
   \end{quotation}

   This enables the model to evaluate whether the player's dialogue contains insights related to the narrative triggers associated with Ryno's origin, which is a necessary condition for progressing to the next level or scene of the game. It also allows the model to output the correct format through the demonstrations given in the prompt.
\end{itemize}

\subsection{Game Interface Design}

The game's setting mimics a social networking platform (Figure~\ref{fig:interface}). Our motivation for this design is to address the question of how to allow players to express attitudes toward climate issues in a similar way to their expression in real-world contexts. We reasoned that the social network style allows players to immerse themselves in the game and behave as they typically would, since it looks just like their own daily use. Even as players gradually uncover a profound and unexpected storyline, the interactions facilitating this revelation remain rooted in mundane, everyday experiences akin to chatting on social media.

Also reflected in the game UI is the deliberate game aesthetic that takes players from the ordinary connection to their own lives to the happenings in a strange new land. The social media interface is designed to lead players to perceive the GPT character as another regular individual conversing with them online. Meanwhile, the progressive infusion of fantastical elements within this reality-based setting, such as sudden revelations of memories and evolving story arcs, evokes a sense of the surreal seamlessly infiltrating the real world.

\section{Methods}\label{sec:Methods}

\begin{figure*}[t]
    \centering
    \includegraphics[width=0.93\linewidth]{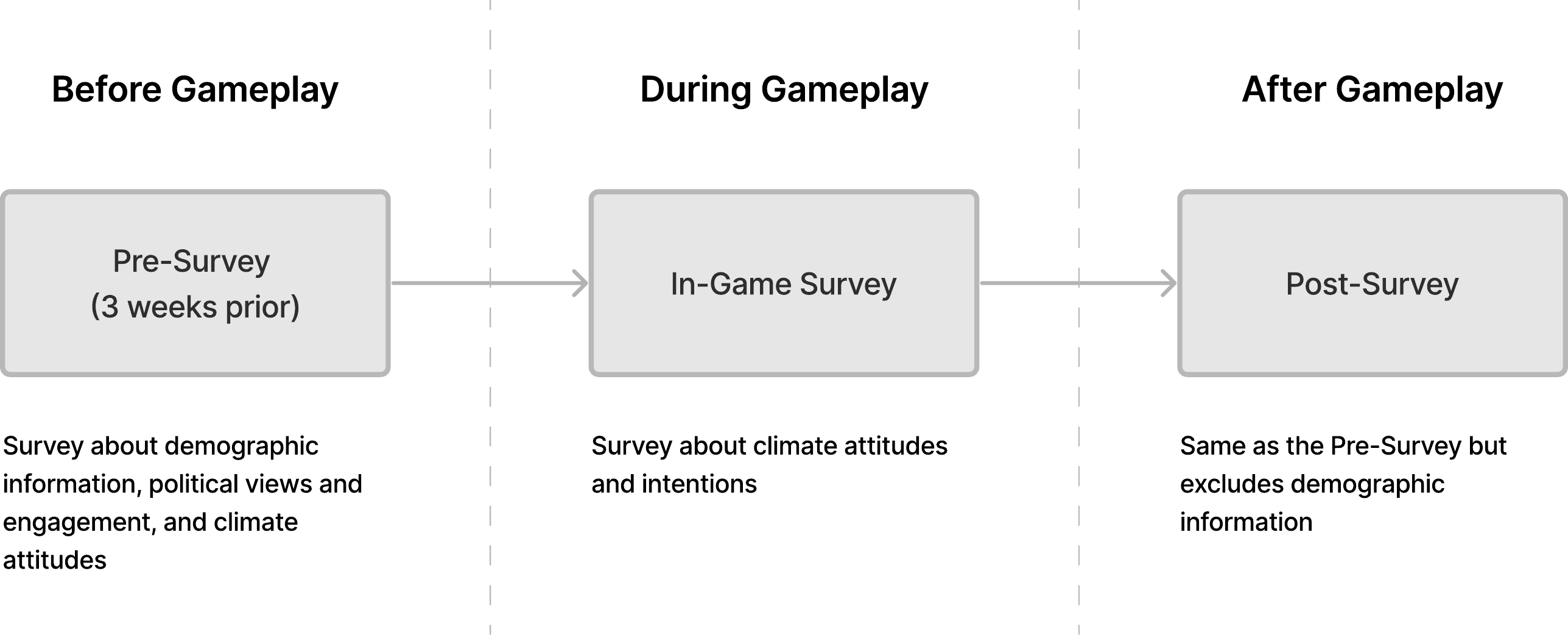}
    \caption{Summary of the research protocol during three distinct periods. Prior to commencing gaming, users are required to fill out a pre-survey that covers demographic information, political perspectives and involvement, as well as climate sentiments. While playing, players participate in an in-game survey that specifically examines their climate opinions and objectives. Following the games, a post-survey is administered that is similar to the pre-survey, with the exception of demographic questions. The purpose of this survey is to evaluate any changes in attitudes or intentions.}
    \label{fig:Study_procedure}
\end{figure*}

\subsection{Recruitment and Participants}

We received replies from 68 participants, with 33 of them completing the entirety of the task and contributing to the data collection. Recruitment was conducted through social networks, specifically Red Book, where we posted a QR code for participation. We opted for Red Book due to its widespread use among our target audience of social media users. Participants received a payment of 50 Hong Kong dollars or Chinese Renminbi upon completion of the game and surveys.
\begin{itemize}
    \item Consent: We ensured that participants provided informed consent for the study. The consent form comprehensively outlined the research's objectives, procedures, potential risks, and benefits. It emphasized the voluntary nature of participation and the participants' right to withdraw at any time.
    \item Anonymity: Participant privacy and anonymity were rigorously protected, and no personally identifiable information was collected. We took precautions to avoid gathering any sensitive data that could be used to identify individuals.
    \item IRB: University ethics review board approved the human subject testing in this project.
    \item Ethic: We adhered to fundamental ethical principles in our study, encompassing the respect of participants' rights, minimization of risks, and the assurance of data confidentiality and security.
\end{itemize}

\begin{table*}[t]
\centering
\caption{Participant Demographics (N=33)}
\label{tab: participants}
\resizebox{0.9\textwidth}{!}{
\begin{tabular}{llllll}
    \toprule
ID  & Gender & Age   & Educational Level           & Occupation Type    & Ethnicity              \\ 
    \midrule
P1  & Male   & 25-34 & Bachelor’s degree           & Self-employed      & Asian/Pacific Islander \\ 
P2  & Male   & 18-24 & Bachelor’s degree           & Employed for wages & Asian/Pacific Islander \\ 
P3  & Male   & 25-34 & Some college                & Self-employed      & White                  \\ 
P4  & Male   & 18-24 & Some college                & Student            & Asian/Pacific Islander \\ 
P5  & Male   & 18-24 & Bachelor’s degree           & Student            & Asian/Pacific Islander \\ 
P6  & Male   & 18-24 & Some college                & Student            & Asian/Pacific Islander \\ 
P7  & Male   & 18-24 & Some college                & Student            & Asian/Pacific Islander \\ 
P8  & Female & 18-24 & Some college                & Student            & Asian/Pacific Islander \\ 
P9  & Male   & 18-24 & Bachelor’s degree           & Employed for wages & Asian/Pacific Islander \\ 
P10 & Male   & 18-24 & Bachelor’s degree           & Student            & Asian/Pacific Islander \\ 
P11 & Male   & 18-24 & Bachelor’s degree           & Student            & Asian/Pacific Islander \\ 
P12 & Male   & 25-34 & Some college                & Self-employed      & Asian/Pacific Islander \\ 
P13 & Female & 25-34 & Bachelor’s degree           & Employed for wages & Asian/Pacific Islander \\ 
P14 & Male   & 18-24 & Bachelor’s degree           & Employed for wages & Asian/Pacific Islander \\ 
P15 & Male   & 18-24 & Bachelor’s degree           & Student            & Asian/Pacific Islander \\ 
P16 & Male   & 18-24 & Some college                & Self-employed      & Asian/Pacific Islander \\ 
P17 & Female & 25-34 & Vocational/associate degree & Employed for wages & Asian/Pacific Islander \\ 
P18 & Male   & 18-24 & Some college                & Student            & Asian/Pacific Islander \\ 
P19 & Female & 18-24 & Some graduate school        & Employed for wages & Asian/Pacific Islander \\ 
P20 & Male   & 18-24 & Some college                & Student            & Asian/Pacific Islander \\ 
P21 & Female & 18-24 & Bachelor’s degree           & Student            & Asian/Pacific Islander \\ 
P22 & Male   & 18-24 & Bachelor’s degree           & Student            & Asian/Pacific Islander \\ 
P23 & Male   & 18-24 & Bachelor’s degree           & Student            & Asian/Pacific Islander \\ 
P24 & Male   & 25-34 & Bachelor’s degree           & Student            & Asian/Pacific Islander \\ 
P25 & Male   & 25-34 & Some college                & Self-employed      & Asian/Pacific Islander \\ 
P26 & Female & 25-34 & Vocational/associate degree & Employed for wages & Asian/Pacific Islander \\ 
P27 & Male   & 18-24 & Some college                & Student            & Asian/Pacific Islander \\ 
P28 & Male   & 25-34 & Vocational/associate degree & Self-employed      & Other                  \\ 
P29 & Male   & 18-24 & Bachelor’s degree           & Student            & Asian/Pacific Islander \\ 
P30 & Female & 18-24 & Some college                & Student            & Asian/Pacific Islander \\ 
P31 & Male   & 25-34 & Some college                & Self-employed      & White                  \\ 
P32 & Male   & 18-24 & Vocational/associate degree & Student            & Asian/Pacific Islander \\ 
P33 & Female & 18-24 & Some college                & Student            & Asian/Pacific Islander \\ 
\bottomrule
\end{tabular}%
}
\end{table*}

\subsection{Survey Design and Data Collocation}

To address the research questions, we conducted a mixed-methods study (Figure~\ref{fig:study_procedure}) involving 33 participants with diverse political and personality backgrounds in China. To explore participants' backgrounds, climate attitudes, and political beliefs, we conducted a pre-survey gathering demographic, political, personality trait, and climate belief data three weeks before the gaming session. In order to prevent participants from preemptively discerning the climate-focused nature of our research and intentionally biasing the results towards pro-climate-leaning responses, the pre-test was given well before the game play (three weeks prior) in order to assure that any subtle information they pick up during the pre-testing would be forgotten by the time they play the game. We also employed a five-point scale and randomized the order of questions.

To address the research questions, we conducted a mixed-methods study involving 33 participants with diverse political and personality backgrounds in China. To explore participants' backgrounds, climate attitudes, and political beliefs, we conducted a pre-survey gathering demographic, political, personality trait, and climate belief data three weeks before the gaming session. In order to prevent participants from preemptively discerning the climate-focused nature of our research and intentionally biasing the results towards pro-climate-leaning responses, the pre-test was given well before the game play (three weeks prior) in order to assure that any subtle information they pick up during the pre-testing would be forgotten by the time they play the game. We also employed a five-point scale and randomized the order of questions.

The first section asked for participants' demographic information, occupation, and ethnicity. As we aim to assess the relationship between personality traits and self-reported environmental attitudes and behaviors, participants were then given an assessment of the Big Five Personality Inventory. We utilized the 50-item set of IPIP Big Five Factor Markers, a pre-validated inventory  ~\cite{goldberg_development_1992}. Study has found that core personality traits, including the Big Five and HEXACO dimensions, are reliable indicators for predicting individual differences in environmental behavior across cultures \cite{brick_unearthing_2016}. Similarly, a study of diverse cultures and languages from five language families supports the universality of the Big Five personality trait structure \cite{mccrae_personality_1997}. Some studies suggest that models predicting environmental attitudes lack a general-level correspondence with behavior \cite{kaiser_environmental_1999}. Given the complexity of environmental behaviors and their dependence on various causal factors, some general theories of environmentalism may not be very useful for changing specific behaviors \cite{stern_new_2000}. Evidence also indicates that early personality, such as childhood temperament, has lasting effects over subsequent decades \cite{block_nursery_2006,caspi_temperamental_1995,slutske_undercontrolled_2012}. Therefore, we can attempt to test a mediation model between core personality, attitudes, values, and beliefs, and then study behavior \cite{brick_unearthing_2016}. The widespread use of the Big Five framework \cite{olver_personality_2003} in various fields is confirmed in questionnaire methodologies, and its close correlation with well-established temperament markers adds further validity \cite{de_fruyt_cloningers_2000,shafer_relation_2001}.

The second survey segment aimed to assess participants' support for democratic norms, evaluation of the political system, and willingness to engage in collective activities, using 18 items~\cite{lei2013political}. Quantitative studies in similar contexts often involve identifying dimensions within political attitudes from selected items and treating each factor as a continuous variable~\cite{norris_cosmopolitan_2009}.

The third section focused on participants' current climate attitudes. Among the 15 items in the Climate Change Attitude Survey, ten were adapted or developed based on literature addressing adult attitudes towards climate change~\cite{leiserowitz_climate_2013}. An additional five items were sourced from work by the Wisconsin Center for Environmental Education~\cite{champeau_r_environmental_1997}. Responses were collected on a Likert-type scale from 1 to 5, where 1 represented "strongly disagree" and 5 denoted "strongly agree." Notably, five negatively worded items were reversed before grouping related items into scales for data analysis. Items measuring intentions were rephrased to capture respondents' intended actions, while belief-oriented items were strategically phrased to elicit judgments about perceived truths in general. We ensured the collection of participants' climate attitudes before the game without them being informed about the game's theme. 


In our text-based adventure game, we integrated an in-game survey presented as dialogue. This survey comprises nine questions, each offering three options. We adjusted the original climate attitude scale to align more closely with the game's contextual design. Participants complete the in-game task after gaining an understanding of the game's background through interactions with ChatGPT-trained character Ryno. The in-game survey primarily aims to assess users' climate attitudes and intentions. All interactions between users and Ryno were documented during this experience.


Following the gameplay phase, we invited all participants to complete a post-survey to assess the connection between self-reported climate attitudes and in-game behaviors after playing the game. In the post-survey, we utilized the same questionnaire as the pre-survey three weeks prior, excluding the demographic section.

\begin{table*}[t]
\centering
\renewcommand{\arraystretch}{1.5} 
\setlength{\tabcolsep}{5pt} 
\caption{Political attitudes coding scale: Response options for survey items on democratic norms (items 1-8), status quo evaluations (items 9–17), and collective action willingness (items 18–19), with codes ranging from 1 (most positive) to 5 (most negative).}
\label{tab:political coding}
\resizebox{0.9\textwidth}{!}{
\begin{tabular}{|c|p{4.5cm}|p{4.5cm}|p{4.5cm}|}
\hline
\textbf{Code} & \textbf{Normative standards (Items 1-8)} & \textbf{Evaluation of the status quo (Items 9–17)} & \textbf{Willingness to participate in collective action (Items 18–19)} \\ 
\hline
1 & Very support the norms of democracy & Very positive evaluation & Lack of willingness \\ \hline
2 & Support the norms of democracy & Positive evaluation & Somewhat lack of willingness \\ \hline
3 & Don't know & Don't know & Don't know \\ \hline
4 & Not support the norms of democracy & Negative evaluation & Demonstrate some willingness \\ \hline
5 & Very not support the norms of democracy & Very negative evaluation & Demonstrate willingness \\ \hline
\end{tabular}}
\end{table*}

\subsection{Data Analysis}
In terms of participants' demographic backgrounds, we coded the data for the 33 participants to facilitate subsequent analyses of the correlation between their basic information such as age, gender, and their attitudes towards climate change. The summary information of participants within each category is illustrated in table 1.

  From the pre-survey, in-game behavior, and post-survey, we analyzed the following:
  
  \begin{itemize}
     \item (a) Correlations between participants' self-reported climate attitudes and in-game behavior.
     \item (b) Relationship between participants' personality, background characteristics, and their self-reported climate attitudes.
     \item (c) Relationship between participants' personality, background characteristics, and their in-game behavior.
  \end{itemize}

We collected data from 33 participants and computed the average climate attitude scores for pre-survey, in-game behavior, and post-survey. Subsequently, we performed Spearman correlation analyses to examine the relationship between participants' self-reported climate attitudes before and after gameplay and their in-game behavior. In the in-game survey, questions are labeled from "IngameQ1" to "IngameQ9," while in the pre- and post-survey datasets, questions are denoted from "PreQ1" to "PreQ14." Both the pre-survey (indicating pre-game climate attitudes) and post-survey (indicating post-game climate attitudes) datasets consisted of 15 related questions about climate attitudes (see Appendix B).

To analyze the relationship between participants' background, personality traits, and self-reported climate attitudes, we utilized the average climate attitude scores based on personality traits and conducted Spearman correlation analyses.

Regarding participants' political backgrounds, we incorporated 19 questions into both pre- and post-surveys to assess their climate attitudes (see Appendix C). Participants' political attitudes were categorized into "Democracy Enthusiasm," "Evaluation of the Status Quo," and "Willingness to Participate in Collective Action." The coding for their political attitudes is outlined in Table~\ref{tab:political coding}. We used Spearman correlation analyses on the whole set of data from before the survey to look at the link between the participants' average scores in each area of political attitudes, their average self-reported climate attitudes, and how they behaved during the game. Then we did the same thing for the complete post-survey dataset.

\section{Results}\label{sec:Results}

\subsection{In-Game Behavior and Self-Reported Climate Attitudes} 

\begin{figure*}[h]
        \centering
        \includegraphics[width=1\linewidth]{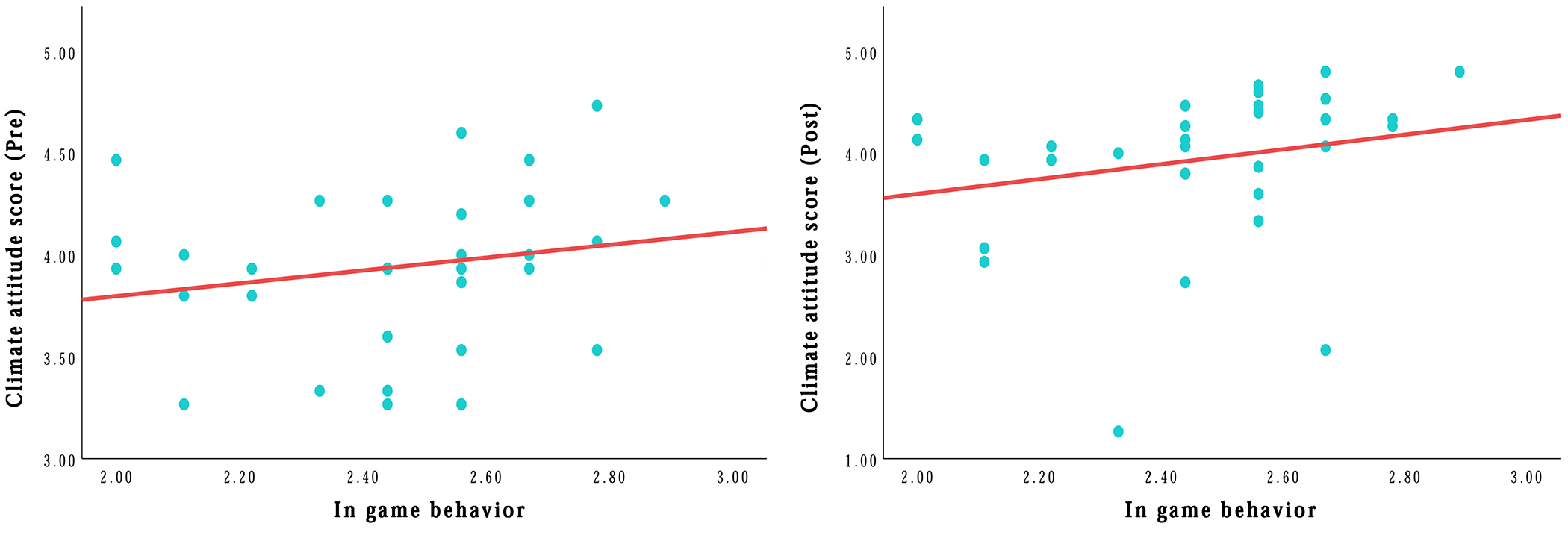}
        \caption{Scatter plots correlating in-game climate attitudes with pre- and post-gameplay assessments. The left plot shows attitudes before gameplay, and the right plot shows attitudes after gameplay, both compared to in-game attitudes. The least squares fit line in both indicates a positive correlation.}
        \label{fig:scatter plot}
    \end{figure*}

Although the goal of our game was assess the attitudes of participants in regards to climate change, we noted interesting effects that our intervention had on participant climate attitudes before and after game play, so we explored both the assessment and attitude change dimensions in our analyses. First, we compared scatter plots of users' climate attitudes averaged across a set of related questions before and during the game with those after the game. This showed a link between how users behaved in the game and how they said they felt about climate change (Figure~\ref{fig:scatter plot}). The trend line illustrates the overall relationship between variables. Participants exhibited a positive trend between climate attitudes in both pre- and post-survey and their in-game behavior. 

\setlength{\intextsep}{20pt} 

While pre-test results (Spearman's R = 0.276, p = 0.120) exhibited no correlation with in-game scores, a significant positive relationship emerged between post-test outcomes and in-game performance (R = 0.390, p =.025). The results, analyzed using Spearman correlation, revealed consistently positive correlations. This implies that individuals achieving higher scores in the game tend to demonstrate elevated post-test scores, despite the pre-test and post-test being identical.

In summary, the analysis indicates an alignment between participants' self-reported attitudes and in-game behavior, a coherence that persists after gameplay. This suggests that game behaviors provide a meaningful metric for assessing attitudes.

\subsection{Personal Traits and Self-reported Climate Change Attitudes.}

To assess the impact of climate change interventions on distinct demographic groups, we partitioned the personality scale in the pre-survey dataset into five variables of interest based on the Big Five Personality Inventory (Table~\ref{tab:my-table}). Subsequently, we calculated participants' average climate attitude scores in the pre-survey, in-game survey, and post-survey, treating these as three additional variables. A Spearman correlation analysis was then conducted on these five variables.


\begin{table*}[t]
\centering
\caption{Spearman correlation analysis between participants' personality traits and climate attitudes (N=33). Notably, 'Openness to Experience' exhibits a significant positive correlation with climate attitudes in both pre- and post-gameplay assessments.}
\label{tab:my-table}
\renewcommand{\arraystretch}{1.5} 
\setlength{\tabcolsep}{10pt} 
\resizebox{0.9\textwidth}{!}{
\begin{tabular}{|l|l|l|l|l|l|}
\hline
\textbf{Personality Trait  }                   & \textbf{Mean}                & \textbf{correlation}& \textbf{INGAME} & \textbf{PRE} & \textbf{POST} \\ \hline
Conscientiousness                     & 3.62                  & R     & .096   & .284  & -.017          \\ \hline
Neuroticism                           & 3.04                  & R     & .112   & .309  & .092 \\ \hline
Openness to Experience                & 3.58                  & R     & .105   & \textbf{.520**} & \textbf{.358*} \\ \hline
Agreeableness                         & 3.78                  & R     & .191   & \textbf{.350*} & .066 \\ \hline
Extraversion                          & 3.40                  & R     & -.028  & .219  & .017 \\ \hline
\multicolumn{6}{|l|}{N=33, R=correlation}                          \\\hline
\end{tabular}}
\end{table*}

A correlation emerged between climate attitudes and the trait of “Agreeableness” (R=0.350, p=.046), suggesting that individuals with a more amiable and empathetic disposition hold more positive beliefs and intentions regarding climate change. Additionally, a consistent correlation was observed between the “Openness” trait in the Big Five personality model and both pretest (R=0.520, p=.002) and post test (R=0.358,p=.041) climate attitude results. This suggests that individuals who are more responsible and open to new experiences tend to exhibit more positive climate attitudes, highlighting the pivotal role of openness in fostering pro-climate attitudes. However, no significant correlation was found between climate attitudes and the traits of “Neuroticism,” “Conscientiousness,” and “Extraversion,” suggesting that factors like individuals' level of emotional stability do not correlate with their climate attitudes.

Overall, the results indicate that “Openness to Experience” and "Agreeableness” serve as parameters for anticipating attitudes and offer insights into designing interventions for different groups. For instance, it suggests the potential use of more implicit influence for individuals with negative climate attitudes.

\subsection{Political Briefs}

We employed Spearman correlation analysis to investigate the relationship between participants' political attitudes and their attitudes toward climate change, encompassing pre-survey and post-survey climate attitude scores (Table~\ref{tab:Political side}).

\begin{table*}[t]
\centering
\caption{Political Elements (Before and After the Gameplay) - Spearman Correlation Analysis (N=33)}
\label{tab:Political side}
\renewcommand{\arraystretch}{1.5} 
\setlength{\tabcolsep}{10pt} 
\resizebox{0.9\textwidth}{!}{
\begin{tabular}{|l|l|l|l|}
\hline
\textbf{Political Element} & \textbf{Correlation} & \textbf{INGAME} & \textbf{PRE} \\ \hline
Support the norms of democracy (Pre) & R & .159 & \textbf{.431*} \\ \hline
Evaluation of the status quo (Pre) & R & -.106 & -.254 \\ \hline
Willingness to participate in collective action (Pre) & R & -.068 & .266 \\ \hline

\textbf{Political Element} & \textbf{Correlation} & \textbf{INGAME} & \textbf{POST} \\ \hline
Support the norms of democracy (Post) & R & .133 & \textbf{.597**} \\ \hline
Evaluation of the status quo (Post) & R & \textbf{-.373*} & \textbf{-.656**} \\ \hline
Willingness to participate in collective action (Post) & R & .206 & \textbf{.353*} \\ \hline
\multicolumn{4}{|l|}{N=33, R=correlation} \\ \hline
\end{tabular}}
\end{table*}

We found that the participants' enthusiasm for democratic values before gameplay significantly correlated with their self-reported climate attitudes in the pre-survey (R=0.431, p=.012), indicating a high consistency between enthusiasm for democracy and positive climate attitudes. However,in the pre-survey, we did not find significant associations between participants' scores on the "Evaluation of the Status Quo" and "Willingness to Participate in Collective Action" indicators with their in-game behavior and self-reported climate attitudes.

Similarly,after the game play,participants' scores on the "Support the Norms of Democracy" indicator correlated significantly (R=0.597, p=.000) with participants' climate attitude scores. In the post-survey, the "Evaluation of the Status Quo" indicator showed significant negative correlations with in-game behavior (R=-0.373, p=.033) and self-reported climate attitudes (R=-0.656, p=.000). The scores on "Willingness to Participate in Collective Action" correlated positively with participants' self-reported climate attitudes (R=0.353, p=.044).

Participants' political passion had positive associations with their self-reported climate opinions, both before and during the game. This indicates that democratic enthusiasm may serve as an indicator of climate attitudes. The climate sentiments stated by participants after playing the game showed strong relationships with all three variables. Consequently, players' assessments of the existing state of affairs and their inclination to engage in collective action might serve as indicators of their climate attitudes after the simulation. Furthermore, the players' assessments of the existing situation might provide insight into their conduct throughout the game.

In summary, individuals' political attitudes can reflect their climate attitudes and in-game behavior. These correlations offer insights for identifying participants and tailoring game interventions, particularly for those with strong self-reported anti-climate sentiments that could be read out in their actual game performance.

\section{Discussion}\label{sec:Discussion}

Various design issues were observed in this approach of game-based evaluation mixed with GPT-supported exploration. This entails using a world-building game to define prompts for GPT, maintaining a harmonious blend of unrestricted exploration via GPT-assisted discourse, and advancing the game through the use of trigger words. The findings of our research indicate that there was no notable link between attitudes towards climate change in the game and the pre-test scores conducted three weeks before the testing. However, a substantial and positive correlation was established between attitudes in the game and attitudes after playing the game. This implies that the players' sentiments about climate change right after playing the game are more similar to their conduct inside the game. Furthermore, we discovered noteworthy associations between the climatic opinions of participants and certain personality qualities, such as Agreeableness and Openness. Ultimately, we observed that individuals with democratic tendencies often had more favorable perspectives on climate change. In the following text, we analyze the consequences that arose from our investigation.

\subsection{Future Climate Scenario and Its Effect on Climate Attitude Assimilation}

Although our focus was on assessment of climate attitudes, the assessed results provided interesting look beyond our expectation about how participant attitudes were assimilated during and after game play. The results demonstrating the correlation between three conditions of climate change attitudes revealed notable findings. Firstly, the significant positive correlation between pre- and post-test scores indicates a fundamental stability in participants' attitudes towards climate change during a three-week interval. This also implies that their basic stance on the issue did not undergo drastic changes due to the gameplay. Participants who were either more or less concerned about climate change before the intervention tended to maintain these attitudes afterward. This consistency over time aligns with our design intention, as our game primarily focuses on assessment rather than altering players' attitudes towards climate change. This outcome also affirms the robustness and reliability of the measurement method and the climate change attitude scale used \cite{christensen_climate_2015}.

Secondly, we observed a correlation between the in-game data and the post-test data, but not with the pre-test data taken for the same individuals three weeks earlier. The absence of correlation between the in-game and pre-test groups suggests a shift in player attitudes towards the same question items over time. This indicates that an effective, immersive game experience in our deliberately designed climate scenarios is capable of being perceived and can eventually result in an attitude change. Another possibility is that the transcription of in-game question items, which are derivatives of the original climate attitude survey, may have led to the attitude shift due to a distortion of the original survey.

To further explain this observation, the strong positive correlation found between attitudes during the game and after the test suggests a trend towards increasing similarity in these attitudes. Considering that the pre- and post-tests were conducted under identical conditions, and taking into account the significant positive correlation already identified between them, it can be inferred that the game functions primarily as an assessment tool rather than an interventional instrument. Were this not the case, a more significant difference between pre- and post-test attitudes would likely have been observed. We interpreted the observation as indicative of an assimilating effect: after playing the game, participants' attitudes towards climate issues became more aligned with those presented within the game world. In a limited capacity, the game subtly modifies these attitudes, guiding them within the framework of their original beliefs rather than completely overhauling them.

The game's layout and scenes may be to blame for this assimilation effect, where players become more in line with their attitudes from the game after playing. We depicted a futuristic climate-related scenario in the game in which the environment is devastated. A key point is that we included a familiar yet distant setting—a 'what if' scenario—for players to explore and experience a potential future of environmental devastation in a distant world. The elements of the narratives of games are therefore not entirely foreign. For example, the reasons for climate issues in the game, such as exploitation of natural resources and political chaos, are topics with which modern humans are familiar \cite{young2002institutional}. More importantly, as the game was designed for assessment purposes, we encouraged players to be as honest about themselves as possible. To do so, we utilized a game UI and setting styled after a social network app, reinforcing the idea that 'you are you' when communicating through social media platforms \cite{bailey2020authentic}.

Interestingly, the technique of pre-experiencing the future also plays a crucial role in exploring potential adaptations to speculative climate challenges and in the field of climate psychology \cite{lc2023human,lc2023together}. It is particularly valuable for studying the psychological distance involved in perceiving climate issues\cite{liberman2007psychological}. This concept originates from the gradual nature of climate change, leading to a subjective perception of climate change as a distant, less immediate problem. Such distance negatively impacts public engagement and complicates the understanding of cognitive responses to climate issues \cite{spence2012psychological, lc2022designing}. To mitigate the effect of psychological distance and promote engagement and pro-climate behaviors, environmental psychologists have adopted the Episodic Future Thinking (EFT) approach \cite{bo2019terrible}. This cognitive method of envisioning future scenarios enables individuals to more tangibly grasp the potential consequences of climate change \cite{lee2020using,bulley2016prospection}. 

For the reasons mentioned above, the observed differences in attitude patterns between the pre-game and in-game conditions suggest that our game effectively creates detailed and vivid representations of futuristic scenarios. This facilitates a high degree of immersion, potent enough to influence attitudes. The fact that players' awareness of climate change issues aligns more closely with the in-game perspective after playing indicates that our game design might effectively 'bring the future closer.' While this effect on altering participants' perceptions of climate issues is unanticipated and limited, it appears to be lasting and significant enough to affect participants' attitudes after playing. However, since the primary objective of this study is assessment, further research is crucial to fully understanding the extent and durability of these changes brought about by pre-experiencing the future.

\subsection{Personal Traits and Political Attitudes towards Climate Change Attitudes}

In our analysis of users' personal traits and their attitudes toward climate change, we found that among the Big Five personality traits, "Openness to Experience" and "Agreeableness" exhibited significant positive correlations with participants' self-reported climate change attitudes. Previous work has explored individual differences in climate change attitudes and environmentalism, consistently finding a positive association between openness and environmental intentions \cite{hilbig_tracing_2013,hirsh_personality_2007}. In a previous study investigating the influence of personality traits on attitudes toward climate change, significant correlations were observed between "openness" and both climate belief certainty and climate change risk perception. Our results support this observation. However, in their study, no significant relationships were found among the other four personality traits \cite{rothermich_influence_2021}. While most studies suggest no association between high levels of neuroticism and engagement in pro-environmental attitudes, a more comprehensive investigation reveals that individuals with higher neuroticism scores do not report engaging in more pro-environmental behaviors \cite{holmstrom_influence_2015}. However, they do exhibit higher levels of pro-environmental attitudes.

In the context of Openness, which encompasses imagination, creativity, and receptiveness to ideas, individuals with higher scores in "Openness to Experience" may signal a greater concern for climate issues and a stronger belief in climate change. In a study on personality predictors of consumerism and environmentalism \cite{hirsh_personality_2007}, results showed significant positive associations between the environmentalism component and both Agreeableness and Openness. Our findings align with this, suggesting that Agreeableness and Openness are effective predictors of specific pro-climate attitudes. Additionally, another study \cite{hilbig_tracing_2013} examining the relationship between Openness to Experience and environmental attitudes and behaviors echoed results from multiple samples \cite{hirsh_personality_2010,hirsh_personality_2007,markowitz_profiling_2012}. This study also indicated a positive link between Honesty–Humility and pro-environmental attitudes. In the exploration of predictors revealing the roots of pro-environmental behavior, literature supports a strong argument for the association between Openness and environmentalism \cite{brick_unearthing_2016}. The flexibility and abstract thinking tied to Openness are considered crucial for recognizing long-term environmental consequences, including those related to climate change. The consistently observed positive correlation between "Openness to Experience" and environmental attitudes and behaviors in previous research is reaffirmed and supported by our findings. Our analysis strengthens the credibility of these prior works. Our findings underscore the significance of personality in gauging individuals' climate change attitudes. Among the Big Five traits, "Openness to Experience" emerges as a particularly potent indicator, demonstrating the strongest association with positive climate change attitudes compared to other traits.

We discovered intriguing results regarding the relationship between individuals' political attitudes and their climate change perspectives. Specifically, a fervor for democracy emerged as a significant indicator with a positive correlation to participants' self-reported climate change attitudes. This finding aligns with previous research that has highlighted the highly politicized nature of climate change as an issue. 
An association between the trait of "openness" and political views was mentioned \cite{rothermich_influence_2021}, but the exploration into this association was not further elaborated. According to a study, political orientation and attitudes toward climate change are both influenced by moral intuitions and self-reported political orientation \cite{dawson_will_2012}. An econometric analysis supports this, revealing that conservative, non-green political identification in the US and Germany is linked to less support for publicly financed climate policies \cite{ziegler_political_2017}. However, low-cost climate-friendly behaviors are unaffected by political affiliation \cite{tobler_addressing_2012}. Similarly, even when controlling for political ideology, an association persists between attitudes toward climate change and behaviors during COVID-19 \cite{latkin_association_2022}. This implies a connection between climate change attitudes and COVID-19 behaviors that is not solely driven by political ideology or trust in scientific information. While our results show a significant correlation between political attitudes and climate change attitudes,future research should delve deeper into this relationship using more complex analyses.

Following the gameplay, our results revealed a significant negative correlation between participants' "evaluation of the status quo" and their in-game behavior. Previous researchers found that people's assessment of the current state is contradictory to their environmental concerns, implying that caring for the environment involves rejecting the status quo \cite{brick_unearthing_2016}. This aligns with the logic in our results. If participants perceive the current situation as harmful to the environment, a negative evaluation of the status quo indicates a more positive attitude toward climate change. Our results also indicate a positive correlation between "willingness to participate in collective action" and participants' self-reported climate attitude change after the game. This is also logically consistent, as previous research implies that individuals with a strong sense of internal control are more likely to believe that their actions can bring about change \cite{kollmuss_mind_2002,moser_making_2004}. Conversely, people who lack confidence in their ability to change the status quo and think that change can only come from powerful others are less likely to engage in ecological behavior because they believe their actions are ineffective. 

By customizing interventions based on these predictors, the game could potentially evoke more nuanced responses and behavioral changes, with the expectation that participants' attitudes may shift more positively in alignment with the targeted predictor variables. This categorization would aid in designing personalized game controls for various types of pro-climate or anti-climate attitudes. For instance, if an individual exhibits high levels of "Openness to Experience" and "Agreeableness" with positive correlations to pro-climate attitudes, the game could emphasize the creative and collaborative aspects of environmental stewardship, fostering a sense of shared responsibility. As a result, we may expect that those people will be more likely to adopt sustainable behaviors and express increased concern for climate issues.For those strongly influenced by political attitudes, interventions might involve framing climate actions in alignment with democratic values. By emphasizing the collective impact of environmentally friendly choices, we may see participants more inclined to view climate action as a social responsibility rather than just a personal choice. Another possible way to intervene is by having ChatGPT automatically assess participants' in-game dialogues and provide different information. For individuals satisfied with the status quo, the game can present more environmental issues and challenges, emphasizing the negative impacts of the status quo. This may lead to participants experiencing a stronger sense of the urgency of environmental issues, resulting in positive climate attitudes. Regarding the "willingness to participate in collective action," personalized game elements can emphasize collaborative game scenarios. We can set triggers in advance to let the chatbot propose cooperative tasks and require players to work together to achieve environmental changes. However, these hypotheses need testing in actual play experiments, and further research is needed to assess the persistence and specific effects of this intervention.

\subsection{Implementing Free Dialogue System within Storytelling}

Video game players often have limited control over conversational interactions with NPCs in games. This limitation can be attributed to two primary factors: First, the dialogue systems in games lack support for natural language input. Secondly, the in-game dialogue, in its minimal form, is fundamentally a list with selectable items, each designed with specific conversational functions. One of the most common dialogue systems is the "dialogue wheel." For example, the Dragon Age and Mass Effect series feature a dialogue wheel where options on the right progress the conversation and those on the left allow for requests for more information, repetition, or clarification of previously mentioned points \cite{bizzocchi2012mass}. Other games like "Dusk of the Gods" and "Elder Scrolls IV" enable players to delve deeper into the dialogue by clicking pre-designed keywords \cite{rennick2021improving}. For example, players could select input (“huh?”) to get the NPC to repeat what they just said. Broadening this approach to allow players more control over both the pace and depth of conversations could enhance their sense of agency in these interactions and offer a solution to avoid re-listening to dialogue unnecessarily \cite{day2017agency}.

However, these systems have not markedly diverged from the traditional format; they remain essentially list-based, failing to facilitate genuine human-to-human conversation. While technological constraints play a significant role, there is also a compelling argument for game designers to be cautious about introducing a truly unrestricted dialogue system. This stems from a fundamental tension in video game development: the aspiration to craft rich, believable narratives versus the necessity to keep the volume of content within manageable limits \cite{rennick2021improving,nur2023emergent}.

To understand this tension, we can trace back to the narrative arc. For example, a three-act structure divides a story into three parts (acts), often called the Setup, the Confrontation, and the Resolution \cite{dancyger2012alternative}. This structure, vital in narrative fiction and first popularized in screenplay writing, also forms the foundation for storytelling in various media, including game narratives \cite{sweet2022diverging,mittell2006narrative}. The issue is that the structure of storytelling itself already draws a line. A so-called story means it must reach its Resolution; even if it doesn't follow the three-act structure, it still has to come to an end \cite{isbouts2012storytelling}. From this perspective, a deliberately pre-designed list of dialogue options is necessary, as it helps preserve the story structure. If dialogue were unlimited, it could lead to an array of plot possibilities, potentially undermining this structure and ultimately leading to the loss of the story itself.

Thus, a key design challenge in implementing free dialogue in storytelling involves enabling natural language input without compromising the story's structural integrity. Our study directly addresses this challenge, aiming to evaluate attitudes towards climate change through scenarios crafted using conversational storytelling powered by GPT. Our approach to resolving this issue, as elaborated in the game design methods section, could offer insightful guidance for developing free dialogue systems in game storytelling. Here, we distill the core reasoning and implications behind our design for future game designers and researchers.

We considered the design and establishment of the game's goal to play a vital role. We set our main game goal, which is to help a stranger recover lost memories, at the very beginning. This strategic arrangement serves a dual purpose: it not only motivates players to engage in dialogue but also, importantly, allows for trial and error. Players can freely type their responses, guided by the pre-established theme of the dialogue. In essence, when players deviate from the goal, they are fully aware that their conversation is becoming irrelevant.

Moreover, through our design process, we recognized the importance of aligning game-level architecture with the narrative arc to facilitate free dialogue in storytelling. As previously discussed, we integrated the main game goal with progressive-level goals. This approach resonates with the concept of dramatic suspense, which posits that a plot typically progresses to pose a series of questions \cite{ryan2008interactive}. For instance, a classic suspense question might be, “Will the hero beat the devil?” In our game, the design of goals echos this format, such as “Does Ryno recall the reason for the climate devastation?” That is to say, players engaging freely in dialogue while simultaneously solving puzzles step by step precisely illustrates how storytelling is constructed. Our approach also aligns with ludonarrative theory, which posits a direct and immediate conflict between the demands of a story and the demands of a game. It suggests that game mechanic design itself is an integral part of narrative design \cite{jenkins2004game}.

\section{Limitation}\label{sec:Limitation}

One limitation is that the majority of our participants were of Asian or Pacific Islander ethnicity, indicating a significant sampling bias. This suggests that our research findings may lack sufficient representativeness and may not encompass the viewpoints and responses of other racial or ethnic groups. Additionally, our study included only participants in the 18–34 age group, with no representation from individuals aged 35 and older. This limitation restricts our understanding of reactions across different age groups, as age can have a significant impact on attitudes towards climate change. Furthermore, over half of our participants were students, with limited representation from other occupational groups. Most participants held undergraduate degrees, and there was no participation from individuals with doctoral, legal, or medical degrees. This implies that we did not encompass highly educated groups, which may possess different cognitive and attitudinal perspectives.

Regarding game development, we are cognizant of the complexities involved in formulating in-game questions. During the design phase, we adhere to the principles of Game-based Assessment Design. Our objective is to integrate assessments within our games in a manner that is both seamless and well-executed, ensuring that they complement the game's narrative and maintain a sense of cohesiveness \cite{jaramillo-alcazar_mobile_2017}. To this end, we modify the original questions to align with the narrative context of the game world. 

However, this strategy comes with some risks. The biggest is that the real information in the question items might not be transcribed correctly, which can happen because of human error \cite{kuklinski2000reconsidering}. To mitigate this, we considered using prompt engineering to generate in-game question items. The advantage of employing this technology lies not only in avoiding direct bias but also in facilitating a stealth assessment design \cite{shute2011stealth}. Nonetheless, as previously mentioned, the management of AI-generated content has not yet achieved full semantic precision especially towards safe and relevant topics like climate issue \cite{white_prompt_2023}. Moreover, in the context of experimental methodology, maintaining consistency in the measurement scales used throughout the study is crucial. To prevent the negative impact that could result in data loss, we ultimately decided to implement manually created but unified in-game questions.

Our game was primarily designed in English, which may have led to decreased immersion and comprehension among participants who were not well-versed in the language. Additionally, occasional interface instability during interactions with characters driven by machine learning occasionally resulted in premature game exits, adversely affecting the user experience. Despite these limitations, we are confident that the results and methodologies presented in this paper will be valuable for future research.

\section{Future Work}\label{sec:Future Work}

There are several directions for future work. The current version of our game focuses on assessment, enabling researchers to observe people's attitudes toward climate change within a designed scenario. However, as an exploratory work, the current version only supports participants' immersion in one climate disaster scene. A more advanced assessment tool that we plan to implement in the future is a game that allows for dynamic adjustments of the climate issue theme and the level of disaster. 
We consider such a dynamic assessment system to be technically feasible. The potential benefits of this system include a more controllable test environment that can meet various research demands, and the ability to observe different attitude patterns under granular levels of climate scenario simulation, which can deepen our understanding of specific climate impacts.

Additionally, we aim to deepen our understanding of our design and improve it. To achieve this, we plan to conduct a user study focusing on usability and game experience. We hope to gain insights into player engagement levels during gameplay, the accessibility and comprehensibility of the game content, and the naturalness of the GPT dialogue flows. By acquiring this user data, we can better iterate our design to enhance its assessment function.

While the current study relies on quantitative results to explore attitude trends among different conditions, we are also interested in implementing a dialogue recording mechanism in our game to capture qualitative data from players. This approach aims to target more nuanced aspects of players' attitudes towards climate change, which quantitative methods may not reveal. For instance, we might observe frequent response patterns in conversations between the chatbot and the player at certain stages or identify specific tones of language that players use to express their views on climate issues.

Furthermore, the results of our study have demonstrated that our design is a promising advancement in episodic future thinking (EFT) techniques, potentially leading to alternative applications such as functioning as an educational and intervention tool. As noted in previous EFT literature, effective EFT, particularly in aiding the comprehension of the climate crisis's consequences, requires projecting oneself into future events using concrete mental representations. Our design has made considerable progress in this area, especially in its ability to create future scenarios.

To align more closely with this direction, revisions to certain aspects of our game design methodology are necessary. Currently, our game allows players to pre-experience scenarios as themselves, but it frames climate devastation as an event in an alternate future rather than a future that they inhabit. We are considering a redesign of the narrative to integrate the player as a future resident. This approach aims at a more direct and subject projection of the player in the scene and is technically feasible by reconfiguring the corpus data to better meet these narrative requirements.

Finally, our attempts to integrate the GPT-driven chatbot system within interactive narratives have illuminated a research path for designers and researchers in the game study field. We've pinpointed a crucial challenge: preserving the integrity of the narrative arc while facilitating natural input systems and free-flowing dialogue in storytelling. Disrupting the story structure can lead to the loss of the story. We've realized the importance of strategically arranging game goals to provide essential motivation and dialogue themes that enable free dialogue. Additionally, we emphasize the significance of aligning level design with the storytelling arc in game development. This harmonization allows storytelling to be carried out through dialogue between the player and the chatbot.

\section{Conclusion}\label{sec:Conclusion}
Our study introduces a new game-based method for assessing attitudes toward climate change. This method employs a text-based adventure game set in a futuristic climate scenario, where players interact with a GPT-powered chatbot. The results from examining players' attitudes across in-game and out-of-game conditions reveal an assimilation effect. Notably, players' awareness of climate change issues tends to align more closely with what they perceived in the game after playing it. Further analysis suggests that personality traits such as 'Openness' and 'Agreeableness', along with a tendency toward democratic values, may serve as indicators of pro-climate attitudes. Through our design, we emphasize the critical role of strategically placed game objectives and highlight the importance of maintaining the integrity of the narrative arc. This approach facilitates natural input systems and free-flowing dialogue in storytelling.

\newpage
\onecolumn
\section{Appendix}\label{sec:Appendix}


\appendix 
\section{In-game and Pre-survey Question List}
\label{app:A}

\renewcommand{\arraystretch}{2}

\begin{table*}[ht]
    \setlength{\tabcolsep}{9pt} 
    \resizebox{0.95\textwidth}{!}{
    \begin{tabular}{|p{4.5cm}|p{7.5cm}|p{1.5cm}|}
        \hline
        \textbf{Original} & \textbf{In-game} & \textbf{Category} \\
        \hline
        I believe our climate is changing. & "Now that you've learned about our world, It's hard to ignore how different our world has become, isn't it?" & Belief \\
        \hline
        I am concerned about global climate change. & "For us, the natives... it's quite a transition. I'm wondering, how does this new reality make you feel?" & Belief \\
        \hline
        Global climate change will impact our environment in the next 10 years. & "As time goes on, it's becoming quite apparent. Looking ahead, do you think our surroundings will continue to deteriorate at this rate in the next decade?" & Belief \\
        \hline
        Global climate change will impact future generations. & "Predicting is always hard, isn't it? How do you feel about the kind of world we're leaving for the younger generations? It's quite a heavy thought." & Belief \\
        \hline
        The actions of individuals can make a positive difference in global climate change. & "Many inhabitants on this planet believe the situation is a deadlock. But do you think we as individuals can still shift the tide?" & Belief \\
        \hline
        I can do my part to make the world a better place for future generations. & "Considering the ripple effect of our actions, do you ever wonder if we're helpless, or can we still take steps to improve our situation? It's an encouraging thought, that even amidst all this, we can do something to improve the world for future generations." & Belief \\
        \hline
        Things I do have no effect on the quality of the environment. & "Sigh… It can be overwhelming, and one might wonder if individual actions really have an impact on the overall state of things. I mean, when we're talking about people versus our environment..." & Intention \\
        \hline
        It is a waste of time to work to solve environmental problems. & "But it's a good perspective to keep, acknowledging that collective efforts matter. Do you ever feel like efforts to improve the situation are futile?" & Intention \\
        \hline
        Knowing about environmental problems and issues is important to me. & "So, do you think it's important to stay informed about the issues affecting our world? We should all have the right to know, after all. How important do you think it is to stay clued up on what's happening around us?" & Intention \\
        \hline
    \end{tabular}}
\end{table*}

\renewcommand{\arraystretch}{1}

\newpage

\section{Pre-survey and Post-survey Question List}
\label{app:B}

\renewcommand{\arraystretch}{1.4}

\begin{table*}[ht]
    \centering
    \setlength{\tabcolsep}{10pt} 
    \resizebox{0.9\textwidth}{!}{
    \begin{tabular}{|c|p{6cm}|c|}
        \hline
        \textbf{Question Number} & \textbf{Statement} & \textbf{Category} \\
        \hline
        1 & I believe our climate is changing. & Belief \\
        \hline
        2 & I am concerned about global climate change. & Belief \\
        \hline
        3 & I believe there is evidence of global climate change. & Belief \\
        \hline
        4 & Global climate change will impact our environment in the next 10 years. & Belief \\
        \hline
        5 & Global climate change will impact future generations. & Belief \\
        \hline
        6 & The actions of individuals can make a positive difference in global climate change. & Belief \\
        \hline
        7 & Human activities cause global climate change. & Belief \\
        \hline
        8 & Climate change has a negative effect on our lives. & Belief \\
        \hline
        9 & We cannot do anything to stop global climate change. & Belief \\
        \hline
        10 & I can do my part to make the world a better place for future generations. & Intentions \\
        \hline
        11 & Knowing about environmental problems and issues is important to me. & Belief \\
        \hline
        12 & I think most of the concerns about environmental problems have been exaggerated. & Intentions \\
        \hline
        13 & Things I do have no effect on the quality of the environment. & Intentions \\
        \hline
        14 & It is a waste of time to work to solve environmental problems. & Intentions \\
        \hline
        15 & There is not much I can do that will help solve environmental problems. & Intentions \\
        \hline
    \end{tabular}}
\end{table*}

\renewcommand{\arraystretch}{1}

\newpage

\section{Political Attitude Question List}
\label{app:C}

\renewcommand{\arraystretch}{2}

\begin{table*}[ht]
    \centering
    \setlength{\tabcolsep}{10pt} 
    \resizebox{0.95\textwidth}{!}{
    \resizebox{0.95\textwidth}{!}{
    \begin{tabular}{|c|p{7cm}|c|}
        \hline
        \textbf{Question Number} & \textbf{Statement} & \textbf{Category} \\
        \hline
        1 & Having a strong leader who does not have to bother with parliament and elections. & Democracy Enthusiasm \\
        \hline
        2 & Having experts, not government, to make decisions. & Democracy Enthusiasm. \\
        \hline
        3 & Having the army rule. & Democracy Enthusiasm \\
        \hline
        4 & Having a democratic political system. & Democracy Enthusiasm \\
        \hline
        5 & Choosing leaders in free elections is an essential characteristic of democracy. & Democracy Enthusiasm \\
        \hline
        6 & Civil rights protection against oppression is an essential characteristic of democracy. & Democracy Enthusiasm \\
        \hline
        7 & It is so important for me to live in a country that is governed democratically. & Democracy Enthusiasm \\
        \hline
        8 & We should let protecting freedom of speech be among the top aims of this country. & Democracy Enthusiasm \\
        \hline
        9 & I trust in the press. & Evaluation of the Status Quo \\
        \hline
        10 & I trust in TV. & Evaluation of the Status Quo \\
        \hline
        11 & I trust in police. & Evaluation of the Status Quo \\
        \hline
        12 & I trust in the courts. & Evaluation of the Status Quo \\
        \hline
        13 & I trust in my central government. & Evaluation of the Status Quo \\
        \hline
        14 & I trust in my local government. & Evaluation of the Status Quo \\
        \hline
        15 & I trust in political parties. & Evaluation of the Status Quo \\
        \hline
        16 & My country is governed democratically today. & Evaluation of the Status Quo \\
        \hline
        17 & Human rights are respected in my country. & Evaluation of the Status Quo \\
        \hline
        18 & I am willing to sign a petition. & Willingness to Participate in Collective Action \\
        \hline
        19 & I am willing to join in boycotts. & Willingness to Participate in Collective Action \\
        \hline
    \end{tabular}}}
\end{table*}

\renewcommand{\arraystretch}{1}


\newpage
\twocolumn
\bibliographystyle{ACM-Reference-Format}
\bibliography{references}


\end{document}